\documentclass[a4paper,11pt]{article}

\usepackage{jcappub} 

\renewcommand{\thesection}{\arabic{section}}

\usepackage{hyperref}
\usepackage{amsmath}
\usepackage{amssymb}
\usepackage{subfigure}
\usepackage{graphicx}
\usepackage{color}
\usepackage[usenames,dvipsnames,svgnames,table]{xcolor}

\renewcommand{\thetable}{\arabic{table}}
\numberwithin{equation}{section}

\title{\boldmath Modified Einstein's gravity to probe the sub- and super-Chandrasekhar limiting mass white dwarfs: a new perspective to unify under- and over-luminous type Ia supernovae}

\author{Surajit Kalita}
\author[1]{and Banibrata Mukhopadhyay\note{Corresponding author.}}

\affiliation{Department of Physics, Indian Institute of Science, Bangalore 560012, India}

\emailAdd{surajitk@iisc.ac.in}
\emailAdd{bm@iisc.ac.in}


\abstract{Type Ia supernovae (SNeIa), used as one of the standard candles in astrophysics, are believed to form when the mass of the white dwarf approaches Chandrasekhar mass limit. However, observations in last few decades detected some peculiar SNeIa, which are predicted to be originating from white dwarfs of mass much less than the Chandrasekhar mass limit or much higher than it. Although the unification of these two sub-classes of SNeIa was attempted earlier by our group, in this work, we, for the first time, explain this phenomenon in terms of just one property of the white dwarf which is its central density. Thereby we do not vary the fundamental parameters of the underlying gravity model in the contrary to the earlier attempt. We effectively consider higher order corrections to the Starobinsky-$f(R)$ gravity model to reveal the unification. We show that the limiting mass of a white dwarf is $\sim M_\odot$ for central density $\rho_c \sim 1.4\times10^8$ g/cc, while it is $\sim 2.8M_\odot$ for $\rho_c \sim1.6\times 10^{10}$ g/cc under the same model parameters. We further confirm that these models are viable with respect to the solar system test. This perhaps enlightens very strongly the long standing puzzle lying with the predicted variation of progenitor mass in SNeIa.}

\begin{document}
\maketitle
\flushbottom

\section{Introduction}
Einstein's general theory of relativity is an incredible theory to explain various astrophysical phenomena and early universe cosmology. It provides an immense understanding of the physics of various compact objects, e.g. black holes, neutron stars \citep{compact} etc. However some recent observations in cosmology and also compact objects question the complete validity of general theory of relativity in extremely high density regions \citep{0004-637X-517-2-565,1538-3881-116-3-1009,0004-637X-607-2-665}. Starobinsky first used modified theory of Einstein's gravity to explain some of these problems in cosmology \citep{1979ZhPmR..30..719S}. Eventually many theories and models have been proposed to explain various epochs of the universe \citep{2008PhRvD..77d6009C,2015JCAP...06..032A,2015IJMPD..2441002C} and physical properties of various astrophysical objects \citep{PhysRevD.77.024040,Carvalho2017,PhysRevD.97.084026}. One of the theories which is commonly used is the $f(R)$ gravity, first proposed by Buchdahl \citep{1970MNRAS.150....1B}. Eventually it has been used to study neutron stars \citep{2013JCAP...12..040A,2014PhRvD..89j3509A,2015JCAP...01..001A,2017CQGra..34t5008A,2016IJMPS..4160130A,PhysRevD.89.064019,1475-7516-2014-06-003,2011JCAP...07..020A} and quark stars \citep{ASTASHENOK2015160}, by various models of $f(R)$ gravity. In case of white dwarfs, because of the reason that it has a small compactness factor $\sim 10^{-4}<<1$ \citep{compact}, general relativistic treatment is generally not very important as opposed to the case of neutron stars.

If a progenitor star has mass $\lesssim 8M_\odot$, at the end of its lifetime, it becomes a white dwarf. The inward gravitational force of the white dwarf is balanced by the force due to outward electron degeneracy pressure. If a white dwarf has a binary partner, it starts pulling matter out from the partner due to its high gravity resulting in the increase of mass of the white dwarf. When it gains sufficient amount of matter, at a certain mass, known as Chandrasekhar limit \citep{1931ApJ....74...81C} (currently accepted value $\sim 1.4M_\odot$ for a carbon-oxygen white dwarf), the pressure can no longer balance the inward gravitational pull and it burns out to produce type Ia supernova (SNeIa) with extremely high luminosity. All the SNeIa have similar peak intensity due to their same/similar progenitor mass. Hence they are used as one of the standard candles in astrophysics to measure luminosity distances of various cosmological objects \citep{1987ApJ...323..140L,1997Sci...276.1378N}. Nevertheless, recent observations have detected several peculiar over-luminous SNeIa such as SN 2003fg, SN 2006gz, SN 2007if, SN 2009dc \citep{2006Natur.443..308H,2010ApJ...713.1073S}. These over-luminous SNeIa are believed to be originating from white dwarfs of mass as high as 2.8$M_\odot$ \citep{2006Natur.443..308H}. On the other hand, some other type of peculiar SNeIa such as SN 1991bg, SN 1997cn, SN 1998de, SN 1999by, SN 2005bl \citep{1992AJ....104.1543F,1997MNRAS.284..151M,1998AJ....116.2431T,2001PASP..113..308M,2004ApJ...613.1120G,2008MNRAS.385...75T} has been detected with extremely low luminosity which are inferred to have been originating from white dwarfs of mass as low as 0.5$M_\odot$. In both the scenaries, Chandrasekhar mass limit is well violated. Das and Mukhopadhyay \citep{2015JCAP...05..045D} (hereinafter Paper I), for the first time, argued that this can well be explained by means of a $f(R)$ gravity. Also they were able to link the sub- and super-Chandrasekhar limiting mass white dwarfs and underlying SNeIa by means of the Starobinsky model of $f(R)$ gravity. They used $f(R)=R+\alpha R^2$, where $R$ is the Ricci scalar and $\alpha$ is the parameter of the model. They argued that if $\alpha>0$, it resulted in sub-Chandrasekhar limiting mass white dwarfs and while for $\alpha<0$, it resulted in super-Chandrasekhar white dwarfs. Later appropriate constraints were put in the parameter space, restricted by the observations of various models of $f(R)$ gravity \citep{2017JCAP...10..004B}. Eventually $f(R,T)=R+2\lambda T$ model was used to describe similar physics of the white dwarfs \citep{Carvalho2017}.

Based on the above discussion of $f(R)$ gravity, it is well noted that in all the models, parameters are appropriately chosen and varied to describe the physical properties of the white dwarfs. But varying the fundamental parameters of a model is not a good idea in physics in particular to explain the physics/astrophysics of the same system. In this paper, we show, for the first time, that if we fix the values of the parameters appropriately, just by varying some properties of the white dwarf, we can achieve sub- and super-Chandrasekhar limiting mass white dwarfs. In section \ref{mod gravity}, we introduce the $f(R)$ model based on which we solve the problem. In section \ref{solution procedure}, we discuss very briefly about the solution procedure. Subsequently in section \ref{Result}, we discuss our results as well as the validity of the model through solar system tests. Finally we end with conclusions in section \ref{conclusion}.

\section{Basic equations in the modified gravity model}\label{mod gravity}
Einstein-Hilbert action provides the field equation in general relativity. With the metric signature (-,+,+,+), in 4 dimensions, it is given by \citep{MTW}
\begin{align}\label{Einstein Hilbert Action}
S=\int\Big[\frac{c^4}{16 \pi G}R+\mathcal{L}_M\Big]\sqrt{-g}d^4x,
\end{align}
where $c$ is the speed of light, $G$ the Newton's gravitational constant, $\mathcal{L}_M$ the Lagrangian of the matter field and $g=\det(g_{\mu\nu})$ is the determinant of the metric $g_{\mu\nu}$. Throughout our discussion, we assume the following definitions.
Affine connection $\Gamma^{\alpha}_{\beta\gamma}$ is defined by
\begin{equation}
\Gamma^{\alpha}_{\beta\gamma} = \frac{1}{2}g^{\alpha\sigma}(g_{\beta\sigma,\gamma}+g_{\gamma\sigma,\beta}-g_{\beta\gamma,\sigma}),
\end{equation}
where comma (`$,$') denotes the partial derivative and Greek indices $\alpha, \beta, \dots$ run from 0 to 3. Now Riemann tensor $R_{\alpha\beta\gamma\sigma}$, which gives an idea about curvature of the space-time, is given by
\begin{equation}
R_{\alpha\beta\gamma\sigma}=g_{\alpha \delta}R^{\delta}_{\beta\gamma\sigma}=g_{\alpha \delta}(\partial_\gamma\Gamma^\delta_{\beta \sigma}-\partial_\sigma\Gamma^\delta_{\beta \gamma}+\Gamma^\delta_{\tau\gamma}\Gamma^\tau_{\beta \sigma}-\Gamma^\delta_{\tau\sigma}\Gamma^\tau_{\beta \gamma}).
\end{equation}
Similarly, Ricci tensor and Ricci scalar are defined as follows,
\begin{align}
R_{\alpha\beta} &= R^{\delta}_{\alpha\delta\beta},\\
R &= g^{\alpha\beta}R_{\alpha\beta}.
\end{align}
Now varying the Einstein-Hilbert action with respect to the metric tensor and equating it to zero with appropriate boundary conditions, we obtain the Einstein's field equation for general relativity, given by
\begin{align}\label{Einstein equation}
G_{\mu\nu} = R_{\mu\nu}- \frac{R}{2}g_{\mu\nu} = \frac{8\pi G}{c^4} T_{\mu\nu} ,
\end{align}
where $T_{\mu\nu}$ is the energy-momentum tensor of the matter field.

In case of $f(R)$ gravity, only modification we have to make is that Ricci scalar $R$ has to be replaced by $f(R)$ in the Einstein-Hilbert action of equation (\ref{Einstein Hilbert Action}) without changing anything in the Lagrangian of the matter field. Therefore the modified Einstein-Hilbert action is given by \citep{2010LRR....13....3D,2017PhR...692....1N}
\begin{align}
S=\int\Big[\frac{c^4}{16 \pi G}f(R)+\mathcal{L}_M\Big]\sqrt{-g}d^4x.
\end{align}
Now if we vary this action with respective to the metric tensor, with appropriate boundary conditions, we have the following modified Einstein equation
\begin{equation}\label{modified equation}
f'(R)G_{\mu \nu}+\frac{1}{2}g_{\mu \nu}[Rf'(R)-f(R)]-(\nabla_\mu \nabla_\nu-g_{\mu \nu}\Box)f'(R)=\frac{8\pi G}{c^4}T_{\mu \nu},
\end{equation}
where $f'(R)$ is derivative of $f(R)$ with respect to $R$, $\Box$ is the d'Alembertian operator given by $\Box=\nabla^\mu\nabla_\mu$ and $\nabla_\mu$ is the covariant derivative defined as $\nabla_\mu A_\nu=\partial_\mu A_\nu-\Gamma^{\lambda}_{\mu\nu}A_{\lambda}$ and $\nabla_\mu A^\nu=\partial_\mu A^\nu+\Gamma^{\nu}_{\mu\lambda}A^{\lambda}$. For $f(R)=R$, it is obvious that equation (\ref{modified equation}) will reduce to the Einstein field equation given in equation (\ref{Einstein equation}).

Starobinsky, in his model, used $f(R)=R+\alpha R^2$ \citep{1980PhLB...91...99S}. But in place of $R^2$, we choose $h(R)$, which is some function of $R$ \citep{2014PhRvD..89j3509A}. Therefore, for $f(R)=R+\alpha h(R)$, modified Einstein equation takes the form
\begin{equation}\label{mod equation}
[1+\alpha h_R(R)]G_{\mu \nu}+\frac{\alpha}{2}g_{\mu \nu}[Rh_R(R)-h(R)]-\alpha(\nabla_\mu \nabla_\nu-g_{\mu \nu}\Box)h_R(R)=\frac{8\pi G}{c^4}T_{\mu \nu},
\end{equation}
where $h_R(R)$ is the partial derivative of $h(R)$ with respect to $R$. It is obvious that if $h(R)=R^2$, this will reduce to the Starobinsky model as described in Paper I \citep{2015JCAP...05..045D}. Note that $R$ is proportional to the density of the star.

Motivated by the study by Astashenok et al. \citep{2014PhRvD..89j3509A}, we choose higher order corrections in the Starobinsky model. In subsequent sections, we first choose $h(R)=R^2(1-\gamma R)$, where $\gamma$ being parameter of the model, which leads to $f(R)=R+\alpha R^2(1-\gamma R)$. Next we consider higher order contributions of $\gamma$, e.g. $f(R)=R+\alpha R^2(1-\gamma R+\frac{1}{2}\gamma^2 R^2-\frac{1}{6}\gamma^3 R^3)$. The motivation for choosing these forms of $f(R)$ is the following. We have seen in Paper I \citep{2015JCAP...05..045D} that for $f(R)=R+\alpha R^2$, positive $\alpha$ leads to sub-Chandrasekhar limiting mass white dwarfs. Therefore the above choice ensures the low central density to reveal similar trend. Because due to appropriate $\alpha$ and $\gamma$, $R^2$ term could be dominating over $\mathcal{O}(R^3)$ and higher order terms. However at high enough central density, super-Chandrasekhar limiting mass white dwarf is revealed when negative terms with $\mathcal{O}(R^3)$ and higher order terms of the series dominate over the preceding positive terms. This is effectively similar to the choice of negative $\alpha$ in paper I \citep{2015JCAP...05..045D}. Finally we choose the infinite series of powers of $\gamma$ resulting in $f(R)=R+\alpha R^2 e^{-\gamma R}$. It is obvious that this reduces to the Starobinsky model at small $\gamma$. Below we show eventually that if we fix the values of $\alpha$ and $\gamma$, white dwarf can easily attain the sub- and super-Chandrasekhar limiting masses just depending on its central density. This idea was stated earlier without detailed exploration \citep{curr}.

\section{Solution Procedure}\label{solution procedure}
To have the interior solution of any star, one has to solve the Tolman-Oppenheimer-Volkoff (TOV) equations with appropriate boundary conditions. Therefore our first aim is to obtain modified TOV equations from the given $f(R)$ model. To start with, we choose a spherically symmetric metric which describes the interior of a non-rotating star. The line element is given by
\begin{align}
ds^2 = -e^{2\phi}c^2dt^2+e^{2\lambda}dr^2+r^2(d\theta^2+\sin^2\theta d\phi^2),
\end{align}
where $\phi$ and $\lambda$ are the functions of radial co-ordinate $r$ only. We assume perfect, static and non-magnetized fluid for which $T_{\mu\nu}$ is given by \citep{2014PhRvD..89j3509A}
\begin{align}
T_{\mu\nu}=\text{diag}(e^{2\phi}\rho c^2,e^{2\lambda}P,r^2P, r^2\sin^2\theta P),
\end{align}
where $\rho$ is the density of matter and $P$ is the pressure of fluid. Now substituting all these relations in the modified Einstein equation (\ref{mod equation}), we obtain the following field equations \citep{2014PhRvD..89j3509A}. First equation is obtained while substituting $\mu=\nu=0$ and second equation is obtained by substituting $\mu=\nu=1$ in equation (\ref{mod equation}), given by
\begin{eqnarray}\label{rho eqn}
-8\pi\rho G/c^2 &=& -r^{-2}+e^{-2\lambda}(1-r\lambda')r^{-2}+\alpha h_R[-r^{-2}+e^{-2\lambda}(1-2r\lambda')r^{-2}]\nonumber \\ &-& \frac{1}{2}\alpha(h-h_RR)+e^{-2\lambda}\alpha[h_R'r^{-1}(2-r\lambda')+h_R'']
\end{eqnarray} and
\begin{eqnarray}\label{P eqn}
8\pi G P/c^4 &=& -r^{-2}+e^{-2\lambda}(1+r\phi')r^{-2}+\alpha h_R[-r^{-2}+e^{-2\lambda}(1+2r\phi')r^{-2}]\nonumber \\ &-& \frac{1}{2}\alpha(h-h_RR)+e^{-2\lambda}\alpha h_R'r^{-1}(2+r\phi'),
\end{eqnarray}
where prime (`$'$') denotes the single partial derivative and double-prime (`$''$') denotes the double partial derivative with respect to $r$.

Assuming the solution in the exterior of a star to be Schwarzschild solution, we have the following relation \citep{2010PhRvD..82f4033C}
\begin{equation}
e^{-2\lambda} = 1-\frac{2GM(r)}{c^2r},
\end{equation}
where $M(r)$ is the mass of the star inside the radius $r$. Moreover, to obtain the modified TOV equations, we adopt the first order perturbative approach. In this method, we assume that $|\alpha R| <<1$, such that second and other higher order terms of $\alpha$ can be neglected. Also in the perturbative approach, all the variables are expanded in terms of $\alpha$ and restricted up to first order only, i.e.
\begin{equation}
\begin{aligned}
M&=M^{(0)}+\alpha M^{(1)},\\
P&=P^{(0)}+\alpha P^{(1)},\\
\rho&=\rho^{(0)}+\alpha \rho^{(1)}.
\end{aligned}
\end{equation}
Now from the conservation of energy-momentum tensor, we have
\begin{equation}\label{conservation}
\nabla_\mu T^{\mu\nu} =0
\implies \frac{dP}{dr} = -(P+\rho c^2)\frac{d\phi}{dr}.
\end{equation}
Combining these relations with equation (\ref{rho eqn}), we obtain
\begin{eqnarray}\label{mod_TOV1}
\frac{dM}{dr} &=& 4\pi r^2\rho -\alpha\Bigg[4\pi r^2 \rho^{(0)}h_R+\frac{c^2}{4G}(h-R h_R)r^2\nonumber \\ &+& \frac{1}{2}h_R'\Big(4\pi r^3 \rho^{(0)}+3M^{(0)}-2\frac{c^2}{G}r\Big)-\frac{c^2}{2G}h_R''r^2\Big(1-\frac{2GM^{(0)}}{c^2r}\Big)\Bigg].
\end{eqnarray}
Similarly combining the perturbed relations with equations (\ref{conservation}) and (\ref{P eqn}), we obtain
\begin{eqnarray}\label{mod_TOV2}
\frac{dP}{dr} &=& -\frac{(P+\rho c^2)}{1-\frac{2GM}{c^2r}}\Bigg[\frac{G}{r^2}\Big(\frac{4\pi r^3P}{c^4}+\frac{M}{c^2}\Big)-\alpha \Bigg(4\pi r h_R P^{(0)} \frac{G}{c^4}+\frac{r}{4}(h-Rh_R)\nonumber \\ &+&h_R'\Big(1-\frac{3G M^{(0)}}{2rc^2}+2\pi P^{(0)} r^2 \frac{G}{c^4}\Big)\Bigg)\Bigg].
\end{eqnarray}
The Ricci scalar appeared in the equation turns out to be only of zeroth order and is given by 
\begin{align}
R\approx R^{(0)} = \frac{8\pi G}{c^4}(\rho^{(0)} c^2-3P^{(0)}).
\end{align}

Equations (\ref{mod_TOV1}) and (\ref{mod_TOV2}) are the modified TOV equations. If $\alpha=0$, they will reduce to the TOV equations in general relativity. Moreover, as mentioned by Wald \citep{wald}, the proper mass of a star is given by
\begin{equation}\label{proper mass}
M_P = 4\pi\int_0^{R_s} \rho(r) r^2 \Bigg[1-\frac{2GM(r)}{c^2r}\Bigg]^{-1/2} dr,
\end{equation}
where $M(r)$ is the mass of the star within a radius $r$, as obtained from the equation (\ref{mod_TOV1}) and $R_s$ is the radius of the star which corresponds to the point where pressure becomes zero.

\subsection{Equation of state and boundary conditions}
\begin{figure}[htp]
\centering
\includegraphics[scale=0.70]{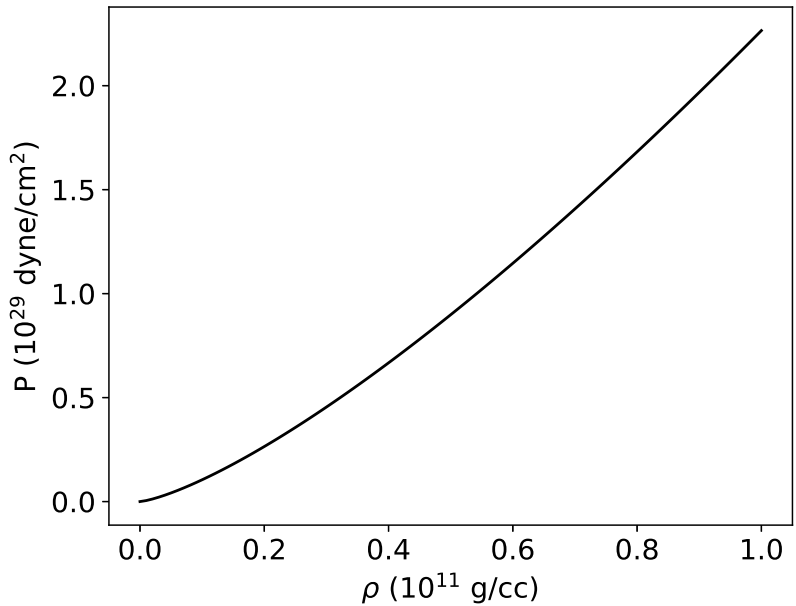}
\caption{Chandrasekhar EoS describing relation between density and pressure.}
\label{Chandra EoS}
\end{figure}

To solve the TOV equations, we have to supply an equation of state (EoS) which relates pressure and density of the system. Since we are considering white dwarfs which contain degenerate electrons, we use Chandrasekhar's EoS, given by \citep{1935MNRAS..95..207C}
\begin{equation}\label{Chandrasekhar EoS}
\begin{aligned}
P &= \frac{\pi m_e^4 c^5}{3 h^3}[x(2x^2-3)\sqrt{x^2+1}+3\sinh^{-1}x],\\
\rho &= \frac{8\pi \mu_e m_H(m_ec)^3}{3h^3}x^3,
\end{aligned}
\end{equation}
where $x = p_F/m_ec$, $p_F$ is the Fermi momentum, $m_e$ the mass of electron, $h$ the Planck's constant, $\mu_e$ the mean molecular weight per electron and $m_H$ the mass of hydrogen atom. For our work, we choose $\mu_e=2$ indicating the carbon-oxygen white dwarf. The Chandrasekhar's EoS of the electron degenerate matter is shown in figure \ref{Chandra EoS}.

Now we have to solve equations (\ref{mod_TOV1}), (\ref{mod_TOV2}) and (\ref{Chandrasekhar EoS}) simultaneously to obtain the interior solution of the star. We use fourth order adaptive Runge-Kutta method to solve these simultaneous equations. The boundary conditions at the center of the star are $M(r=0)=0$ and $\rho(r=0)=\rho_c$. On the other hand, at the surface of the star, we have $\rho(r=R_*)=0$ and $M(r=R_*)=M_*$. We consider the central density $\rho_c$ of the white dwarfs up to a maximum value, well below neutronization threshold. Also we choose the set of values of $\alpha$ and $\gamma$ in such a way that neither it violates perturbation limit nor it violates any conventional physics, i.e. mass should not be zero or negative and they preserve solar system test.

\section{Results}\label{Result}
We consider different models of $f(R)$ based on the higher order corrections of $\gamma$. We show how the results to be varying as we consider more correction terms in the model.

\subsection{$f(R)=R+\alpha R^2(1-\gamma R)$ :}
\begin{figure}[htp]
\centering
\includegraphics[scale=0.5]{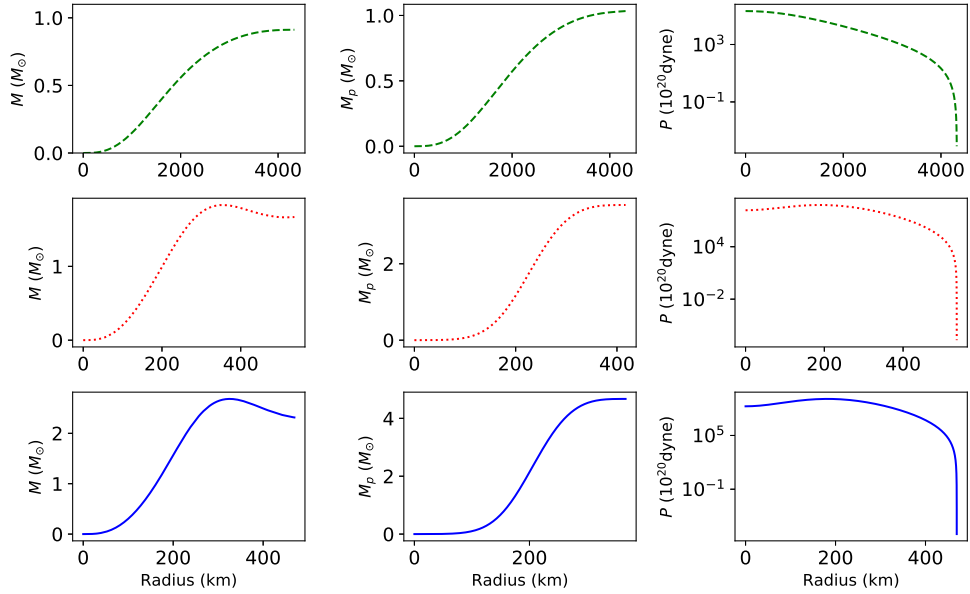}
\caption{The left panels show the variation of mass obtained by solving modified TOV equations (\ref{mod_TOV1}) and (\ref{mod_TOV2}), the middle panels depict the proper mass as defined by the equation (\ref{proper mass}) and the right panels show the variation of pressure with respect to the distance from the center of a white dwarf for different central density. Green-dashed line represents $\rho_c=10^8$ g/cc, red-dotted line represents $\rho_c=1.3\times10^{10}$ g/cc and blue-solid line represents $\rho_c=1.5\times10^{10}$ g/cc.}
\label{mass_density}
\end{figure}

As we have discussed in section \ref{mod gravity}, first order correction to the Starobinsky model can be considered as $f(R)=R+\alpha R^2(1-\gamma R)$. In figure \ref{mass_density}, variations of mass and pressure with respect to the distance from the center of a typical star, which follows this model, are shown. It is seen that due to curvature contribution, the effective mass and pressure of the star change. It is evident from the figure that at high enough central density, the mass of the star has a decreasing trend closer to the surface. This is due to the reason that the Ricci scalar is directly proportional to the density and hence for high density stars, the contribution due to curvature, coming from Ricci scalar, acting against $4\pi r^2 \rho$ in equation (\ref{mod_TOV1}), turns out to be significant.

\begin{figure}[htp]
\centering
\includegraphics[scale=0.70]{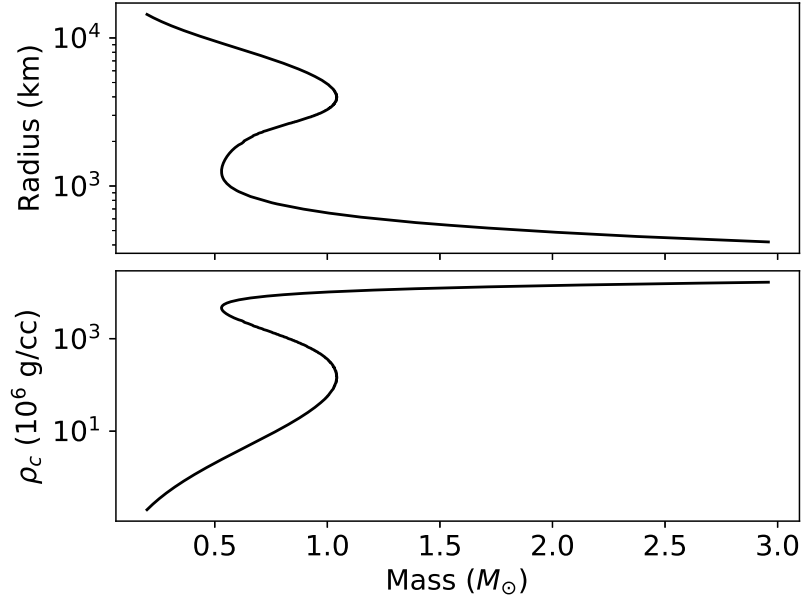}
\caption{Upper panel: The mass-radius relation, Lower panel: The variation of central density with mass of the white dwarf, for $f(R)=R+\alpha R^2(1-\gamma R)$. Here $\alpha=3\times10^{14}$ cm$^2$ and $\gamma=4\times10^{16}$ cm$^2$.}
\label{mass_radiusa3}
\end{figure}

The mass-radius relation as well as the variation of central density with respect to the mass of white dwarf are shown in figure \ref{mass_radiusa3}. Note that the radius and central density are in logarithmic scale. We choose $\alpha=3\times10^{14}$ cm$^2$ and $\gamma=4\times10^{16}$ cm$^2$. At small enough central density, the curve mimics the property of general relativity. Above $\rho_c\sim 1.46\times 10^8$ g/cc, the curve turns back due to dominance of the $\alpha R^2$ term, thereby it mimics the model of Starobinsky gravity as discussed in Paper I \citep{2015JCAP...05..045D}. In this way, it reveals the sub-Chandrasekhar limiting mass white dwarfs, because above $\sim 1.46\times10^8$ g/cc, the stars in the curve are unstable and hence the branch is unstable. In this branch, stellar mass decreases with increasing central density. Since unstable branches are nonphysical, the mass corresponding to this peak of the curve is the limiting mass of white dwarfs which turns out to be sub-Chandrasekhar. We obtain the sub-Chandrasekhar mass limit of white dwarf $\sim 1.04 M_\odot$. Subsequently at further increase of central density, the curve attains another peak at $\rho_c\sim 4.67\times10^9$ g/cc, till which it remains to be unstable branch. Beyond this point, with further increase in central density, again the curve reveals the usual properties of the stellar objects, i.e. increasing mass with increasing central density. This branch of the curve is again physical and stable. As we keep increasing the central density further, the curve goes beyond the Chandrasekhar mass limit. In other words, it approaches to the super-Chandrasekhar mass region. At $\rho_c\sim 1.66\times10^{10}$ g/cc, we have the mass $\sim 2.95 M_\odot$. With the further increase of $\rho_c$, other instabilities related to nuclear physics and general relativity would restrict the mass, revealing actually super-Chandrasekhar limiting mass of white dwarfs (see below).

\subsection{$f(R)=R+\alpha R^2(1-\gamma R+\frac{1}{2}\gamma^2 R^2-\frac{1}{6}\gamma^3 R^3)$ :}
\begin{figure}[htp]
\centering
\includegraphics[scale=0.70]{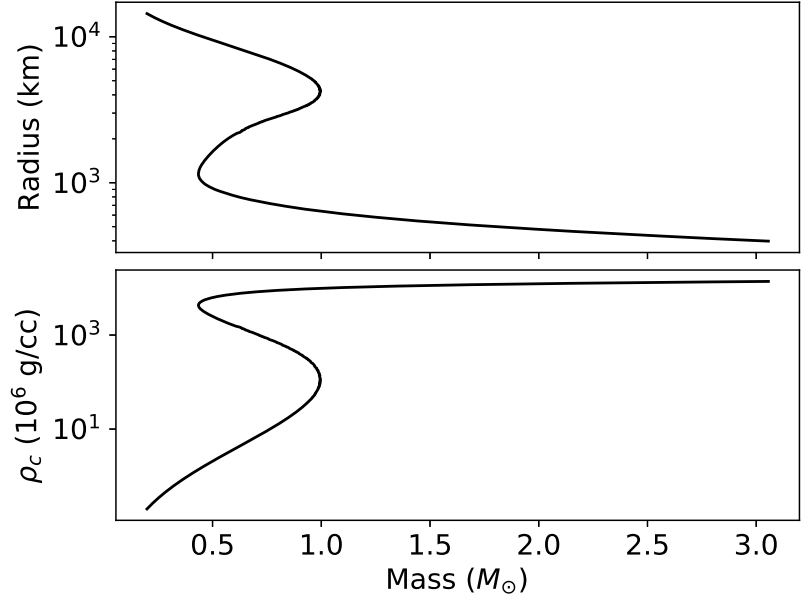}
\caption{Same as figure \ref{mass_radiusa3}, except for $f(R)=R+\alpha R^2(1-\gamma R+\frac{1}{2}\gamma^2 R^2-\frac{1}{6}\gamma^3 R^3)$. Here $\alpha=4\times10^{14}$ cm$^2$ and $\gamma=7\times10^{16}$ cm$^2$.}
\label{mass_radiusa5}
\end{figure}

Once we keep considering higher order terms, the results will (slightly) be modified. Here we choose $\alpha=4\times10^{14}$ cm$^2$, $\gamma=7\times10^{16}$ cm$^2$, and obtain a sub-Chandrasekhar limiting mass white dwarf $\sim 0.99 M_\odot$ at $\rho_c\sim 1.16\times10^8$ g/cc. It attains super-Chandrasekhar mass $\sim 3 M_\odot$ at $\rho_c\sim 1.38\times10^{10}$ g/cc. The variations of mass with radius and central density for this $f(R)$ model are shown in figure \ref{mass_radiusa5}.

\subsection{$f(R)=R+\alpha R^2e^{-\gamma R}$ :}
\begin{figure}[htp]
\centering
\includegraphics[scale=0.70]{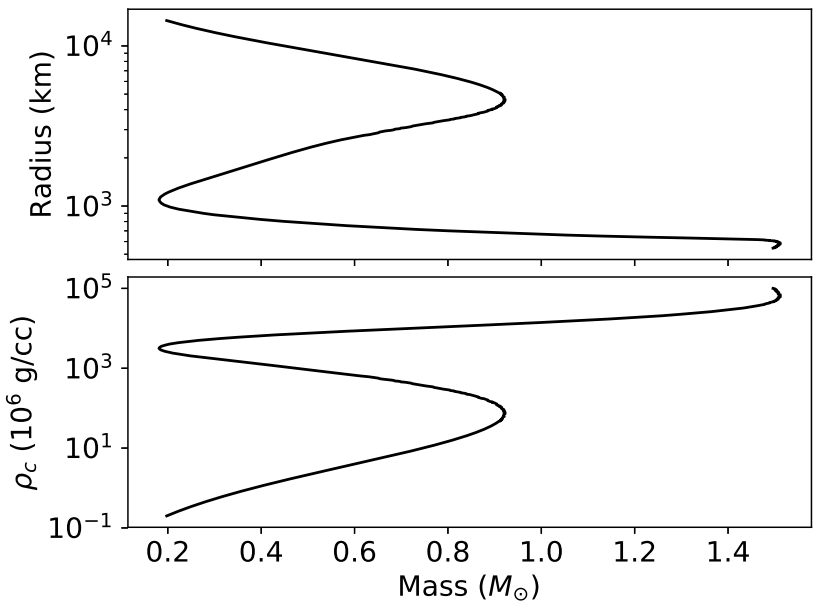}
\caption{Same as figure \ref{mass_radiusa3}, except for $f(R)=R+\alpha R^2e^{-\gamma R}$. Here $\alpha=6\times10^{14}$ cm$^2$ and $\gamma=1.0\times10^{17}$ cm$^2$.}
\label{mass_radius}
\end{figure}

As mentioned in the section \ref{mod gravity}, if we consider all the higher order terms in $\gamma$, it will effectively lead to $f(R)=R+\alpha R^2e^{-\gamma R}$. This reduces to the Starobinsky model when $\gamma$ is small, which in turn reduces to Einstein gravity when $\alpha$ is small. The mass-radius relation and the variation of central density with mass of the white dwarf for this $f(R)$ model are shown in figure \ref{mass_radius}. We choose $\alpha=6\times10^{14}$ cm$^2$ and $\gamma=1.0\times10^{17}$ cm$^2$. Here we attain a sub-Chandrasekhar limiting mass of white dwarf $\sim 0.92 M_\odot$ at $\rho_c\sim 8.9\times10^7$ g/cc. However, at high enough $\rho_c$, the mass-radius curve exhibits third peak, clearly indicating second limiting mass apart from the one at lower $\rho_c$ at the first peak. On the other hand, first order and third order corrections to Starobinsky model do not show any high density peak, hence a clear evidence of limiting mass at high $\rho_c$. Nevertheless, this model reveals marginally super-Chandrasekhar limiting mass $\sim 1.5 M_\odot$ at $\rho_c \sim 10^{11}$ g/cc. Therefore it is expected that at higher order correction of $\gamma$ between third order and exponential, the mass-radius curve will exhibit a third peak and hence limiting mass at significantly super-Chandrasekhar regime.

\subsection{Higher order corrections of $\gamma$ :}
\begin{figure}[htp]
\centering
\includegraphics[scale=0.70]{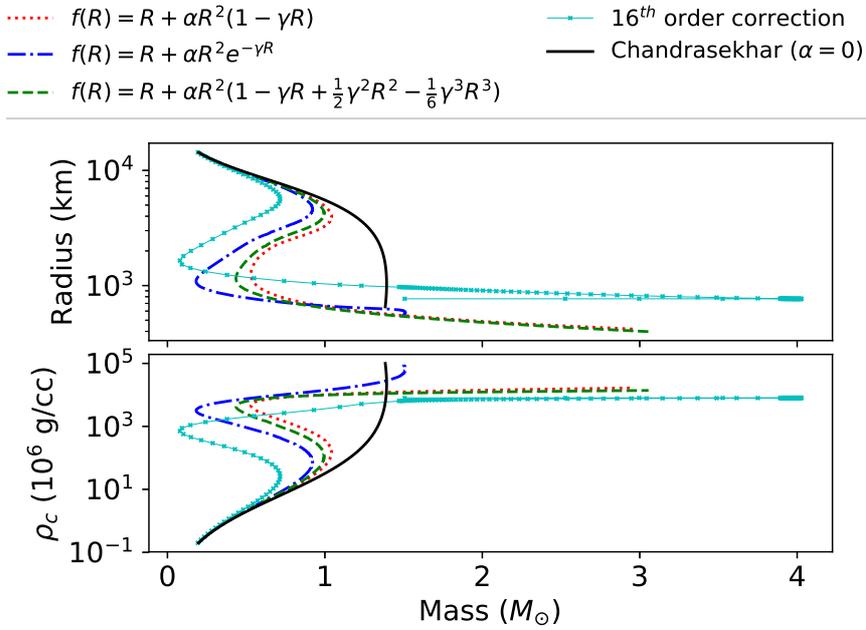}
\caption{Combined result showing the mass-radius relation and the variation of the central density with the mass of white dwarfs for different $f(R)$ models considered here. Note that the radius and central density are in logarithmic scale.}
\label{Combined}
\end{figure}

Now we extend our investigation to the higher order in $\gamma$. Let us consider sixteenth order correction in $\gamma$, with $\alpha=1.5\times10^{15}$ cm$^2$ and $\gamma=4.0\times10^{17}$ cm$^2$. A combined result, illustrating variations of mass with radius and central density of the white dwarfs, for all these models along-with Chandrasekhar's model, is shown in figure \ref{Combined}. It is evident from the cyan solid-point curve for the sixteenth order correction in $\gamma$ that the mass-radius curve can indeed turn back at significantly super-Chandrasekhar mass region hence revealing super-Chandrasekhar limiting mass. Since the exponential model contains all powers of $\gamma$, with alternate positive and negative coefficients, the mass-radius curve turns back at even lower mass, hence revealing lower super-Chandrasekhar limiting mass. Therefore, at appropriate corrections to the Starobinsky model, sub- and super-Chandrasekhar limiting masses, following unstable branches, are possible.

All the above results suggest that we obtain the sub- and super-Chandrasekhar limiting mass white dwarfs by just varying the central density of the star rather varying the parameters of the model. In this way, we can verify the predicted values of sub- and super-Chandrasekhar limiting mass white dwarfs given in various literatures \citep{2006Natur.443..308H,2010ApJ...713.1073S,1992AJ....104.1543F,1997MNRAS.284..151M,1998AJ....116.2431T,2001PASP..113..308M,2004ApJ...613.1120G,2008MNRAS.385...75T}. Moreover it is also evident that the mass-radius relation strictly depends on the curvature contribution, which was also discussed \citep{2016PhRvD..93b3501C} in the context of neutron star.

\subsection{Solar system test}
As discussed by Guo \citep{2014IJMPD..2350036G}, a model is called viable if it passes through the solar system test. He mentioned certain conditions for a model to pass through the solar system test. According to that, an $f(R)$ model of the form $f(R)=R+A(R)$ passes through solar system, if it obeys the following conditions
\begin{equation}\label{solar system eqns}
\begin{aligned}
\Big|\frac{A(R)}{R}\Big|&<<1,\\
|A'(R)|&<<1,\\
RA''(R)&<<1,
\end{aligned}
\end{equation}
where prime (`$'$') denotes the partial derivative with respect to $R$. The numerical values of the left hand sides of these relations, for the $f(R)$ models mentioned above, are given in table \ref{solar system table}. It is well noted that they all satisfy the relations given in equation (\ref{solar system eqns}) and hence these models well pass through the solar system test. Moreover, the final limiting mass or the mass corresponding to the maximum central density for each of these models, are reported in table \ref{solar system table} along-with the mass $M_P$ obtained from equation (\ref{proper mass}). It is evident that $M_P>M$, which is in accordance with Wald \citep{wald}.

\begin{table}[htp]
\centering
\caption{The values of the L.H.S. of relations of equation (\ref{solar system eqns}) and the final limiting proper mass according to the equation (\ref{proper mass}) along-with mass obtained from modified TOV equations (\ref{mod_TOV1}) and (\ref{mod_TOV2}) for super-Chandrasekhar peak, or at the maximum central density considered for these models.}
\label{solar system table}
\resizebox{\columnwidth}{!}{\begin{tabular}{|c|c|c|c|c|c|}
\hline
	$f(R)$ model & $|A(R)/R|_\text{max}$ & $|A'(R)|_\text{max}$ & $[RA''(R)]_\text{max}$&$M (M_\odot)$&$M_P (M_\odot)$\\
\hline\hline
	$R+\alpha R^2(1-\gamma R)$ & $2.19\times 10^{-3}$ & $1.58\times 10^{-2}$ & $1.25\times 10^{-3}$&2.95&5.57\\
\hline
	$R+\alpha R^2(1-\gamma R+\frac{1}{2}\gamma^2 R^2-\frac{1}{6}\gamma^3 R^3)$ & $1.99\times 10^{-3}$ & $1.83\times 10^{-2}$ & $1.18\times 10^{-3}$&3.0&5.29\\
\hline
	$R+\alpha R^2e^{-\gamma R}$ & $2.21\times 10^{-3}$ & $2.77\times 10^{-3}$ & $1.42\times 10^{-3}$&1.5&2.51\\
\hline
\end{tabular}}
\end{table}

\section{Conclusion}\label{conclusion}
The unification of sub- and super-Chandrasekhar limiting mass white dwarfs and under- and over-luminous SNeIa has been done earlier. However, in this work, we, for the first time, have shown that it can be achieved by just varying a single property of the white dwarf, viz. central density, rather than varying the parameters of the model which appears to be ad-hoc. We have chosen simple higher order corrections to the Starobinsky-$f(R)$ gravity model. Based on these models, by fixing the respective set of values of $\alpha$ and $\gamma$ appropriately, we have shown that at low $\rho_c \sim 10^8$ g/cc, the white dwarf has limiting mass well below Chandrasekhar mass limit, producing sub-Chandrasekhar limiting mass white dwarf, while at a higher $\rho_c \gtrsim 10^{10}$ g/cc, the mass of the white dwarf is well above the Chandrasekhar mass limit and revealing super-Chandrasekhar limiting mass white dwarf, depending on the model.

We have used perturbative method to obtain the modified TOV equations. This implies that the value of $\alpha$ should be chosen in such a way that it does not violate the perturbative approximation. The value of $\alpha$ is precisely chosen such that it does not violate the astrophysical constraint given by Gravity Probe B experiment \citep{2010PhRvD..81j4003N} according to which $\alpha \lesssim 5\times10^{15}$ cm$^2$. However, the value of $\gamma$ is chosen in such a way that neither the mass of the white dwarf becomes zero or negative nor it violates the solar system test.

SNeIa are used as one of the standard candles to measure distances in astrophysics and cosmology. The discovery of the peculiar SNeIa prompts us possible modification of the definition of standard candle. At first, people tried to explain over-luminous SNeIa considering a rotating white dwarf. However rotation alone can explain the stable mass up to $\sim 1.8M_\odot$, whereas combining rotation with the magnetic field can explain much more massive white dwarfs. However, none of these effects can explain sub-Chandrasekhar limiting mass white dwarfs and hence the under-luminous SNeIa. Hence some models were proposed to explain this class of white dwarfs, e.g. merger of two sub-Chandrasekhar white dwarfs, reproducing the low power of under-luminous SNeIa. Even though these models were suggested, the major concern still remains in the fact that the need of plenty of different models to explain same physical phenomena. Modification of Einstein's gravity via $f(R)$ gravity using the Starobinsky model, first seemed to solve this long-standing problem to a great extent, although a problem remained that the variations of parameters of the underlying gravity model govern the physics of the system \citep{2015JCAP...05..045D}. To overcome this problem, we, in this work, have suggested higher order corrections to the Starobinsky model. In these various $f(R)$ models, if we fix their respective parameters appropriately, just depending on the central density of the white dwarf, we have achieved sub-Chandrasekhar limiting mass white dwarf at low central density and super-Chandrasekhar limiting mass white dwarf at high central density. Thus we unify the sub- and super-Chandrasekhar limiting mass white dwarfs and thereby probe the unification of peculiar SNeIa, i.e. under- and over-luminous type Ia supernovae.



\end{document}